\documentclass[10pt,letterpaper]{article}
\usepackage{opex3}
\usepackage{color}
\usepackage{threeparttable}

\begin{document}

\title{Wide-field Fourier ptychographic microscopy using laser illumination source}

\author{Jaebum Chung$^{1,*}$, Hangwen Lu$^1$, Xiaoze Ou$^1$, Haojiang Zhou$^1$ and Changhuei Yang$^1$}

\address{$^1$Department of Electrical Engineering, California Institute of Technology, Pasadena, California, 91125, USA }

\email{$^*$jchung@caltech.edu} 



\begin{abstract}
Fourier ptychographic (FP) microscope is a coherent imaging method that can synthesize an image with a higher bandwidth using multiple low-bandwidth images captured at different spatial frequency regions. The method's demand for multiple images drives the need for a brighter illumination scheme and a high-frame-rate camera for a faster acquisition. We report the use of a guided laser beam as an illumination source for an FP microscope. It uses a mirror array and a 2-dimensional scanning Galvo mirror system to provide a sample with plane-wave illuminations at diverse incidence angles. The use of a laser presents speckles in the image capturing process due to reflections between glass surfaces in the system. They appear as slowly varying background fluctuations in the final reconstructed image. We are able to mitigate these artifacts by including a phase image obtained by differential phase contrast (DPC) deconvolution in the FP algorithm. We use a 1-Watt laser configured to provide a collimated beam with 150 mW of power and beam diameter of 1 cm to allow for the total capturing time of 0.96 seconds for 96 raw FP input images in our system, with the camera sensor's frame rate being the bottleneck for speed. We demonstrate a factor of 4.25 resolution improvement using a 0.1 NA objective lens over the full camera field-of-view of 2.7 mm by 1.5 mm. 
\end{abstract}




\section{Introduction}

Fourier ptychographic microscopy (FPM) is a recently developed computational imaging system capable of acquiring the complex and quantitative field distribution of a sample \cite{NatureFPM, OuQuantPhase}. Unlike conventional microscopes that can only image the intensity distribution, FPM's complex sample field contains both its amplitude and phase information. FPM achieves this by a simple modification in sample illumination without the need for a separate reference beam or mechanical movement within the system as in other phase imaging systems. It uses a coherent light source to image different components of the sample's Fourier spectrum, and uses a phase retrieval algorithm to synthesize these images into a high-resolution complex field distribution. Effectively, it can linearly improve the numerical aperture of the imaging lens by the illumination NA. 

There has been various improvements and applications of FPM. Numerical aperture of over 1 for a conventional microscope, usually only achievable by using some immersion medium between objective lens and sample, was realized with a low NA objective and an arrangement of LEDs allowing for steep illumination angles \cite{OuHighNA}. The high-resolution and wide field-of-view (FOV) of FPM showed potential applications in white-blood-cell counting \cite{ChungFPMblood} and resource-limited imaging scenarios \cite{DongFPscope, GuoIllumEngin}. Multiplexed illumination patterns allowed for high resolution high speed phase imaging of unlabeled in-vitro cells \cite{WallerOptica}. An iterative algorithm that reconstructs the aberration of the microscope system simultaneously with the sample spectrum allowed for removal of spatially varying aberrations throughout the microscope's field of view \cite{OuEPRY} and made FPM particularly suitable for imaging samples with uneven surfaces \cite{ChungCTC}. The characterized aberration function further allowed for removing spatially varying aberrations from fluorescence images for even performance across the field of view \cite{ChungBOE}. Insights from FPM carried over to incoherent imaging to improve the resolution of fluorescence images \cite{DongFluorFPM}. There also have been numerous efforts in improving the Fourier ptychographic (FP) reconstruction by adopting more noise-robust algorithms \cite{BianWirtinger, HorstmeyerConvex, BianAdaptiveFPM, DongSparse, YehFPMalgo}. Alternative FPM modalities involving aperture scanning instead of angular illuminations were demonstrated, which allowed for imaging the complex field of a thick specimen \cite{DongFPMB} and estimating optical aberrations \cite{HorstmeyerFPMB}. Imaging a thick specimen with angular illuminations also became possible by employing the first Born approximation in the image formation process \cite{Tian3DFPM, Horstmeyer3DFPM}. 

With wider adoption of FPM for imaging and the desire to image fast dynamics, faster capturing speed is desired. There have been several efforts in this respect. Using LEDs, the required number of captured images can be reduced by optimizing LED illumination arrangements \cite{GuoSamplingPatt, BianContAdaptIllum}, illuminating multiple LEDs of either the same color \cite{WallerMultiplex} or different colors \cite{DongMultiplexColor, DongMultiplexHighDim}, or using better ways of image processing \cite{DongSparse}. Ref. \cite{SoDMD} demonstrated for the first time using a high-power laser beam coupled with a DMD which allowed for shot-noise-limited image capturing process, overcoming the power limitation of LEDs. All these methods address the slow capture issue, but they are not without downsides. By reducing the number of captured images via multiplexing, one increases the shot-noise per individual sub-spectrum of sample. In \cite{SoDMD}, although the power of illumination is easily scalable by using a stronger laser, the on-state mirrors only constitute a small portion of the entire DMD area. The rest of the area which are in the off-state deflects a large portion of the input laser power to a beam dump or in a certain angle which scatters strongly in the optical path and contributes negatively to captured images. Also, the FOV was limited to around 50 ${\rm \mu m}$ by 50 ${\rm \mu m}$ for a 0.04 NA 1.25x objective, which is much smaller than the FOV typically offered by such an objective lens. Another feature overlooked by many FPM illumination schemes implemented so far is the efficient usage of the illumination beam to improve capturing speed: an LED's radiation profile typically follows a Lambertian distribution \cite{NguyenLED} and only a small portion of it actually ends up illuminating the sample; and a DMD only utilizes a tiny fraction of the input laser beam for each sample illumination angle.

Here, we present an FPM setup illuminated by a laser guided by a Galvo mirror and a mirror array. We are able to utilize 15\% of the total laser output power for sample illumination with the proposed collimation setup. The utilization ratio can be increased for a higher-beam-quality laser and finer collimation optics. The benefit of using a mirror array over a condenser lens such as done in \cite{SoDMD} and suggested by \cite{CohenSingleshotPtycho} is that the illuminating wavefront does not suffer from additional aberrations induced by lenses, which can negatively impact FP reconstruction if not properly corrected for. Also, the NA of the condenser lens limits the amount of angular scanning possible for angularly varying illumination before the aberration from the lens becomes severe.

We demonstrate that our system can generate a quantitative image of a sample with 4.25 times the resolution. The coherence of laser leads to speckle artifacts that presents a challenge in our reconstruction process, the majority of which originates from the strong unscattered laser beam reflected between glass surfaces in the optical path. We mitigate the speckle's influence by incorporating into the FP algorithm the sample's phase obtained from differential phase contrast (DPC) deconvolution method \cite{TianDPC} which is a procedure also employed in \cite{WallerOptica} during FPM initialization for better reconstruction of low-frequency phase information. We show that low spatial frequency artifacts due to speckles are effectively removed from the reconstructed phase of the sample. Overall, our laser FPM demonstrates wide FOV and high quality image reconstruction at a high capturing speed.

\section{Principle and algorithm}

\subsection{Principle of FPM}
Our FPM algorithm operates on the principle that the sample to be imaged is very thin \cite{OuHighNA}. This essentially turns it into a two dimensional sample, similar to a thin transparent film with an absorption and phase profile on it. When the sample on the stage of a 4f microscope is perpendicularly illuminated by a light source that is coherent both temporally (i.e. monochromatic) and spatially (i.e. plane wave), the light field transmitted through the sample is Fourier transformed when it passes through the objective lens and arrives at the objective's back-focal plane. The field is then Fourier transformed again as it propagates through the microscope's tube lens to be imaged onto a camera sensor or in a microscopist's eyes. The amount of the sample's detail the microscope can capture is defined by the objective's numerical aperture (${\rm NA}_{\rm obj}$) which physically limits the extent of the sample's Fourier spectrum being transmitted to the camera. Thus, the ${\rm NA}_{\rm obj}$ acts as a low-pass filter in a 4f imaging system with a coherent illumination source.

In the following, we limit our discussion to a one dimensional case. Extending to two dimensions for a thin sample is direct. Under the illumination of the same light source but at an angle $\theta$ with respect to the sample's normal, the field at the sample plane, $\psi_{\rm oblique}(x)$, can be described as:
\begin{equation} \label{eq1}
\psi_{\rm oblique} (x) = \psi_{\rm sample}(x) \exp(jk_0 x \sin \theta)
\end{equation}
where $\psi_{\rm sample}(x)$ is the sample's complex spatial distribution, $x$ is a one dimensional spatial coordinate, and $k_0$ is given by $2 \pi / \lambda$ where $\lambda$ is the illumination wavelength. This field is Fourier transformed by the objective lens, becoming:
\begin{equation} \label{eq2}
\Psi_{\rm oblique} (k) = \int_{- \infty}^{\infty} \! \psi_{\rm sample}(x)  \exp(jk_0 x \sin \theta ) \exp(-jk x) {\rm dx} = \Psi_{\rm sample}(k - k_0 \sin \theta)
\end{equation}
at the objective's back-focal plane, where $\Psi_{\rm oblique}$ and $\Psi_{\rm sample}$ are the Fourier transforms of $\psi_{\rm sample}$ and $\psi_{\rm oblique}$, respectively, and $k$ is a one dimensional coordinate in $k$-space. $\Psi_{\rm sample}(k)$ is shown to be laterally shifted at the objective's back-focal plane by $k_0 \sin \theta$. Because ${\rm NA}_{\rm obj}$ is physically fixed, a different sub-region of $\Psi_{\rm sample}(k)$ is relayed down the imaging system. Thus, we are able to acquire more regions of $\Psi_{\rm sample}(k)$ by capturing many images under varying illumination angles than we would by only capturing one image under a normal illumination. 

Each sub-sampled Fourier spectrum from the objective's back-focal plane is Fourier transformed again by the tube lens, and the field's intensity value is captured by the camera sensor. Due to the loss of phase information in the intensity measurement, the sub-sampled images cannot be directly combined in the Fourier domain. We use a Fourier ptychographic (FP) algorithm, essentially a phase retrieval algorithm, to reconstruct the phase and amplitude of the expanded Fourier spectrum. The algorithm requires the low-passed images to be captured so that each image contains some overlapping region in the Fourier domain \cite{DongSparse}. We allow a 60\% overlap between images, and this redundancy allows for the FP algorithm to infer the missing phase information through an iterative method which is described in the following section.

\subsection{Algorithm}
A high-resolution image of a sample to be reconstructed is initialized with the low-resolution image captured under a normal illumination. To further improve the iteration's starting point and at the same time apply robust phase constraints to facilitate algorithm convergence, we adopt the DPC deconvolution \cite{TianDPC} to obtain the sample's phase image for an additional initialization step. DPC deconvolution is a partially coherent method to achieve the quantitative phase of a sample. It is based on the assumption that the sample's absorption and phase are small such that the sample's complex transmission function, $\psi(x) = \exp(-\mu(x) + j\theta(x))$, can be approximated as \cite{Streibl3D}: $\psi(x) \approx 1 - \mu(x) + j\theta(x)$. Under this condition, performing simple arithmetic operations on the images captured under different illumination angles generates multiple-axis DPC images and the transfer function associated with the sample's phase and the DPC images \cite{TianDPC}. Deconvolving the transfer function from the DPC images results in the quantitative phase image of the sample with the spatial frequency information extending to $2 k_0 {\rm NA}_{\rm obj}$ in $k$-space. Also, the partially coherent nature of DPC method significantly reduces the speckle noise originating from the coherent light source \cite{TianDPC} and thus improves the reconstructed image of our algorithm.

\begin{figure}
\centering\includegraphics[width=13cm]{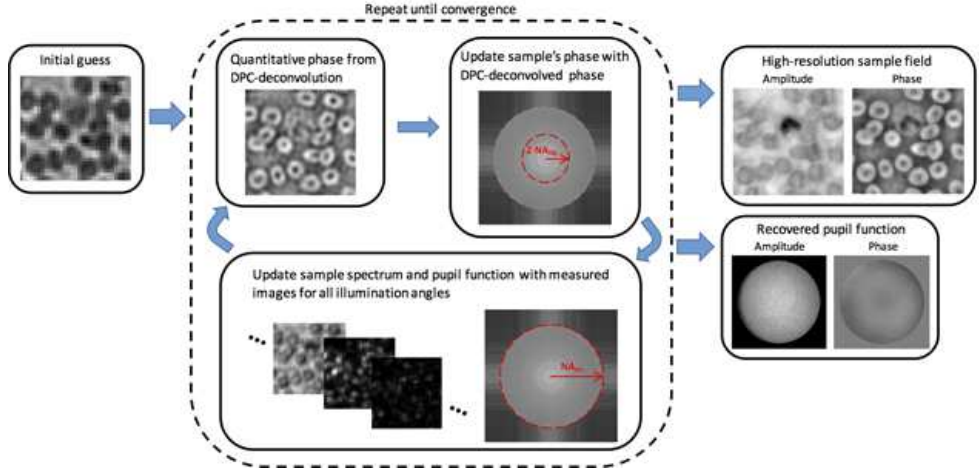}
\caption{\label{fig:algo}Modified FP algorithm to include DPC-generated phase into the iteration. The reconstruction begins with the raw image captured with the illumination from the center mirror element as an initial guess of the sample field. The iteration process starts by forming the sample's quantitative phase image with 2NA resolution by DPC deconvolution with the initial guess of the pupil function. The phase of the sample field up to the 2NA resolution extent is updated. Images captured under varying illuminations are used to update the pupil function and the sample's Fourier spectrum, just as in the original FP algorithm. The updated pupil function is used to generate an updated DPC-deconvolved phase image for the update process, and the iteration process repeats until convergence. In the end, we reconstruct the complex field of the sample and the pupil function.}
\end{figure}

We update the phase of our initial guess with the DPC-deconvolved quantitative phase as follows:
\begin{equation} \label{eq3}
\psi_{2 {\rm NA}}(x) = \left| \mathcal{F} \{ \Psi(k) P_{2{\rm NA}}(k) \} \right| \exp(j  \theta_{\rm DPC})
\end{equation}
where $\Psi (k)$ is the high-resolution Fourier spectrum of a sample, $P_{2 {\rm NA}}$ is the low-pass filter with the spatial frequency extent of $2 k_0 {\rm NA}_{\rm obj}$ in $k$-space, $\mathcal{F}$ is Fourier transform operator, $\theta_{\rm DPC}$ is the quantitative phase obtained from DPC deconvolution, and $\psi_{2 {\rm NA}}$ is the simulated image with its phase updated with $\theta_{\rm DPC}$. Unlike intensity image updates in FP, an update with the phase from DPC deconvolution requires us to use a pupil function extending to $2{\rm NA}_{\rm obj}$ instead of just ${\rm NA}_{\rm obj}$ because the deconvolved phase contains information up to $2{\rm NA}_{\rm obj}$ resolution \cite{TianDPC}. Intensity images captured at different angles are used to reconstruct the high resolution Fourier spectrum and the pupil function of the microscope as done in the original FP algorithm found in \cite{NatureFPM, OuEPRY}. The generation of DPC phase and the update process involving the DPC phase and intensity images constitute one iteration. DPC phase needs to be recalculated at the beginning of each iteration because the pupil function of the microscope changes during pupil function update procedure. The overall algorithm is summarized in Fig. \ref{fig:algo}. For the reconstruction to convergence, we conduct 25 iterations without updating the pupil function, and 15 with, resulting in 40 iterations in total. In the end, we obtain the high-resolution complex field of the sample and the imaging system's pupil function.

\section{Experiments and results}

\begin{figure}
\centering\includegraphics[width=5cm]{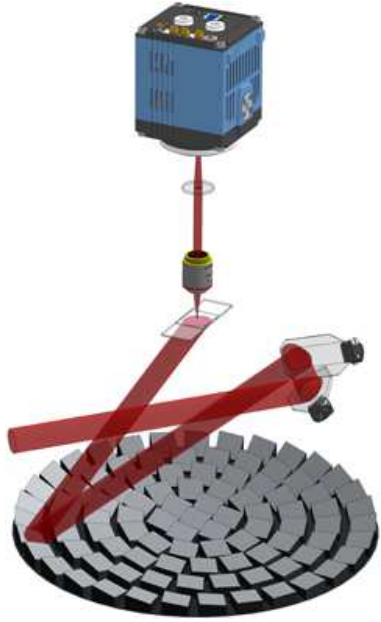}
\caption{\label{setup3D} Experimental setup. It consists of a 4f system with the 2D Galvo mirror system and the mirror array guiding the laser illumination direction. The beam diameter is about 1 cm, covering the entire FOV captured by the camera (2.7 mm by 1.5 mm after magnification). The objective lens has an NA of 0.1 and the total illumination NA is 0.325, resulting in ${\rm NA}_{\rm sys} = 0.425$.}
\end{figure}

\subsection{Setup}
\begin{figure}
\centering\includegraphics[width=10cm]{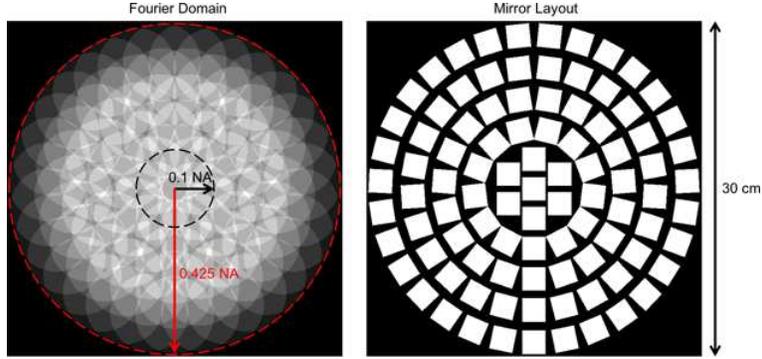}
\caption{\label{fig:layout}The Fourier spectrum region covered by the angularly varying illumination and the layout of the mirror array to achieve the desired coverage. With the objective NA of 0.1 and one normal plane wave illumination, the spatial frequency acquired by the system is delineated by the black circle in the Fourier domain. With varying illumination angles, we can expand the extent of the captured spatial frequency, as indicated by the red circle with the NA of 0.425. The mirror array is 30 cm wide and is placed 40 cm away from the sample plane. Each circular bandpass in the Fourier domain, with its size defined by ${\rm NA}_{\rm obj}$ and its location by the illumination angle provided by each mirror element, has 60\% overlap with the contiguous one.}
\end{figure}

The imaging setup is a 4f system consisting of a 0.1 NA objective lens (Olympus 4x), 200-mm-focal-length tube lens (Thorlabs), and a 16bit sCMOS sensor (PCO.edge 5.5). The sensor has a pixel size of 6.5 ${\rm \mu m}$ and a maximum frame-rate of 100 Hz at 1920x1080 resolution for a global shutter mode. The sensor size limits the available FOV of a sample to be 2.7 mm by 1.5 mm. On the illumination side, 457 nm 1 W laser beam is pinhole-filtered and collimated. After collimation process, the output beam is 1 cm in diameter and 150 mW in power. A set of mirrors guides the beam such that the central part of its Gaussian profile (about 40\% of total output area) is incident on the input of 2D galvo mirror device (GVS 212) for a uniform beam intensity distribution at its output. Galvo then guides the beam to individual mirror elements on the 3D-printed array, as shown in Fig. \ref{setup3D}. Each mirror element is a 19mm x 19mm first-surface mirror attached to a 3D-printed rectangular tower. The tower's top surface is sloped at a certain angle such that the beam from Galvo is reflected towards the sample's location. Thus, the element's spatial location relative to the sample determines the illumination angle of the beam. The mirror array consists of 95 elements arranged to provide illumination angles such that contiguous elements produce 60\% overlap of the sample's spectrum in the Fourier domain, as shown in Fig. \ref{fig:layout}. 
The total illumination NA corresponds to ${\rm NA}_{\rm ti} = 0.325$ with the resulting system NA being ${\rm NA}_{\rm sys} = {\rm NA}_{\rm obj} + {\rm NA}_{\rm ti} = 0.1 + 0.325 = 0.425$, effectively increasing the microscope's NA by a factor of 4.25. 

To achieve the maximum frame rate of the sCMOS sensor in the image capturing process, the exposure time is set to its minimum, at 500 microseconds. The sensor and Galvo are externally triggered every 10 milliseconds, resulting in 0.96 seconds of total capturing time for 95 sample images and 1 dark noise image. Maintaining the same exposure time for all images presents a small challenge: the SNR of the images are drastically different between images captured in the bright-field illumination (${\rm NA}_{\rm illum} < {\rm NA}_{\rm obj}$) and the ones in the dark-field (${\rm NA}_{\rm illum} > {\rm NA}_{\rm obj}$) because the unscattered laser beam comprises the most of the signal from the sample, especially for natural samples such as neurons \cite{HeintzmannObjectSpectrum}. Adjusting the laser intensity for the proper exposure of the bright-field images would result in low signal values in dark field images at the same laser intensity level and camera exposure time. As a result, the dark field images would tend to be more affected by dark noise. To account for this, a neutral density filter is placed on each bright-field illumination mirror elements. This allows for increasing the input laser intensity to obtain higher SNR in dark field images while preventing the bright field images from over-exposure.

\begin{figure}
\centering\includegraphics[width=12cm]{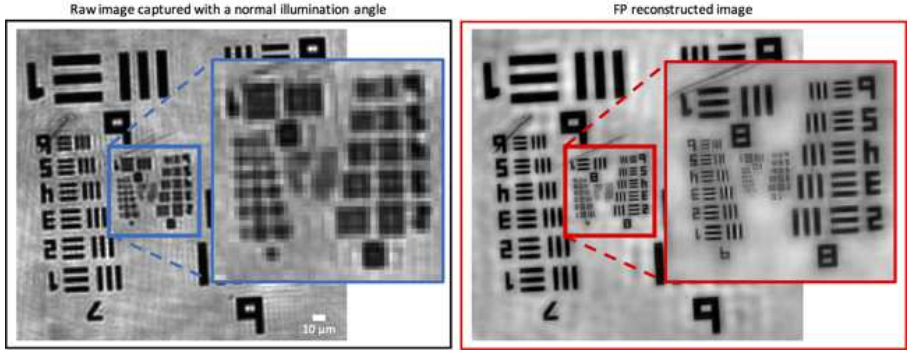}
\caption{\label{fig:USAF}USAF target, before and after FP reconstruction. Under a normal illumination provided by the center mirror element, the theoretical resolution corresponds to $\frac{\lambda}{{\rm NA}_{\rm obj}} = 4.57$ ${\rm \mu}$m periodicity. Up to group 7 element 6 is observed, which has 4.38 ${\rm \mu}$m periodicity, closely matching the prediction. After FP reconstruction, the theoretical resolution is $\frac{\lambda}{{\rm NA}_{\rm sys}} = 1.08$ ${\rm \mu}$m periodicity, closely matching the observed group 9 element 6's periodicity of 1.10 ${\rm \mu}$m.}
\end{figure}

\subsection{Spatial resolution}
We image a positive USAF resolution target printed on a microscope slide to validate our system's resolving power. Under a normal illumination by the laser beam reflected from the center mirror element, the maximum resolvable feature is group 7 element 6, which corresponds to 4.38 ${\rm \mu m}$ periodicity as shown in Fig. \ref{fig:USAF}. We then capture multiple images under varied illumination angles and apply our FP algorithm as in Fig. \ref{fig:algo}. The improved image shows features as small as group 9 element 6, which corresponds to 1.1 ${\rm \mu m}$ periodicity, as shown in Fig. \ref{fig:USAF}. This is in accordance to the theoretical resolution afforded by ${\rm NA}_{\rm sys}$ which can be calculated as: periodicity $=\frac{457{\rm \ nm}}{0.425}=1.08 \ {\rm \mu m}$. In Appendix A, we conduct additional measurements using a Siemens star target for a more accurate quantification of our system's resolution \cite{HorstmeyerStar}.

\begin{figure}[b]
\centering\includegraphics[width=10cm]{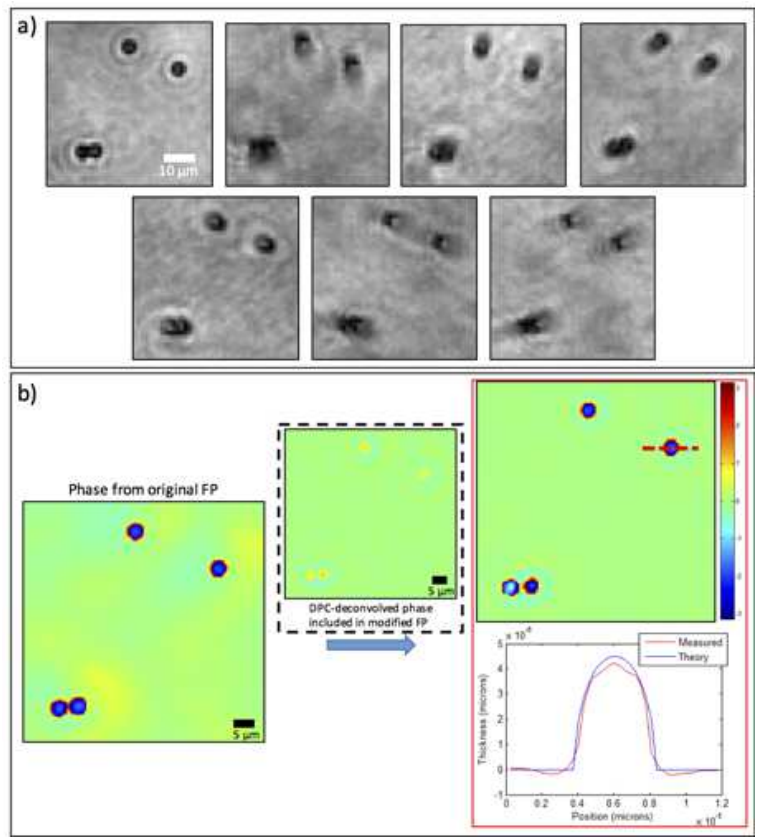}
\caption{\label{fig:quantphase}Images of 4.5-$\rm \mu$m-diameter microspheres sample. (a) Under bright-field illumination (${\rm NA}_{\rm illum} < {\rm NA}_{\rm obj}$), the captured images show fluctuating backgrounds due to the speckle noise from unscattered laser beam reflecting between the reflecting surfaces in the optical path and interfering with itself. (b) Without an additional DPC-deconvolved phase update, the reconstructed phase image shows an uneven background. After the modification, the reconstructed phase is free from the background noise and the resulting phase is also quantitative.}
\end{figure}

\subsection{Quantitative phase}
We use a microscope slide with microspheres to demonstrate quantitative phase imaging of our laser FPM. The sample consists of 4.5-${\rm \mu m}$-diameter polystyrene bead from Polysciences, Inc. (index of refraction $n_s = 1.6119$ @ 457 nm) immersed in oil (index of refraction $n_o = 1.5269$ @ 457 nm). The bead diameters and the indices of refraction for the beads and oil are carefully chosen so that they satisfy the requirement for successful quantitative phase imaging as presented in \cite{OuQuantPhase}. The maximum phase gradient generated by the microbead shouldn't exceed the maximum resolvable spatial frequency of our FPM system. 

In Fig. \ref{fig:quantphase}, we show the importance of including the additional step of DPC-deconvolved phase image update into our algorithm. Fig. \ref{fig:quantphase}(b) shows the FP reconstructed phase image before and after the modification in the algorithm. Background noise in the original FP reconstruction is mainly due to the speckle noise originating from the mostly unscattered laser beam, as seen in Fig. \ref{fig:quantphase}(a), interfering with itself in the optical setup. The noise contributes to the final reconstructed Fourier spectrum as a slowly varying phase signal. By incorporating the DPC deconvolved quantitative phase image in the update scheme, we are able to remove the background noise. This is because the DPC deconvolution is a partially coherent method and thus is immune to speckle noise \cite{TianDPC}. The DPC-deconvolved phase image is free from the influence of the low-fluctuating speckle, as shown in Fig. \ref{fig:quantphase}(b). 

In Fig. \ref{fig:quantphase}(b), one representative micro-bead, as indicated by the dashed line in the updated FP reconstruction phase, is compared with the theoretical bead profile. The reconstructed phase of the bead is unwrapped and converted into the bead thickness using the given refractive indices of oil and beads. The measured bead diameter is 4.24 ${\rm \mu m}$, which is within the 10\% tolerance of the theoretical value of 4.5 ${\rm \mu m}$. 

\begin{figure}
\centering\includegraphics[width=9cm]{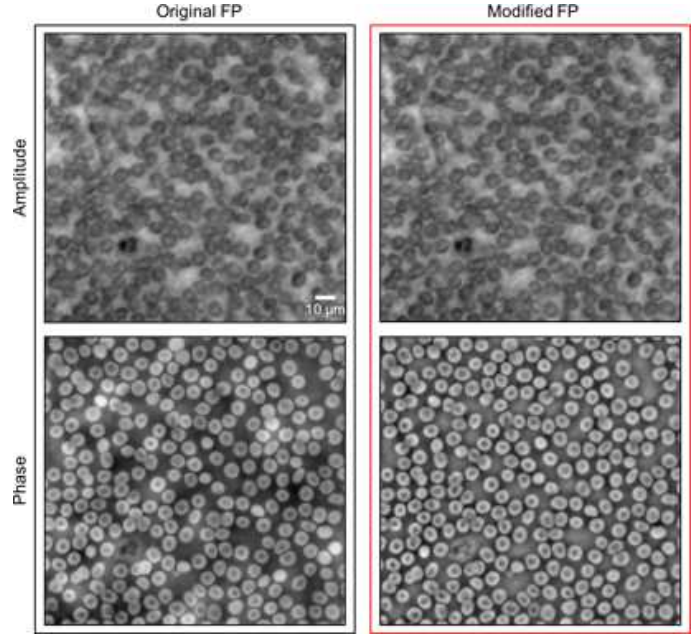}
\caption{\label{fig:blood}Blood smear images, before and after modification in FP algorithm. Without the additional DPC-deconvolved phase update in the reconstruction process, the resulting phase of the sample shows an uneven background signal that also influences the cells' phase amplitude. After the modification, the background is uniform and the red blood cells show similar phase values. Note, the modification has little or no affect on the amplitude image.}
\end{figure}

\subsection{Imaging of biological samples}
We first image a blood smear sample prepared on a microscope slide stained with Hema 3 stain set (Wright-Giemsa). We notice coherence artifacts, i.e. speckles, in the background of the low-resolution captured images. Without the DPC algorithm, the phase image suffers from uneven background signals due to speckles, as shown in Fig. \ref{fig:blood}. After the DPC update, we observe the phase clears up significantly.

\begin{figure}
\centering\includegraphics[width=9cm]{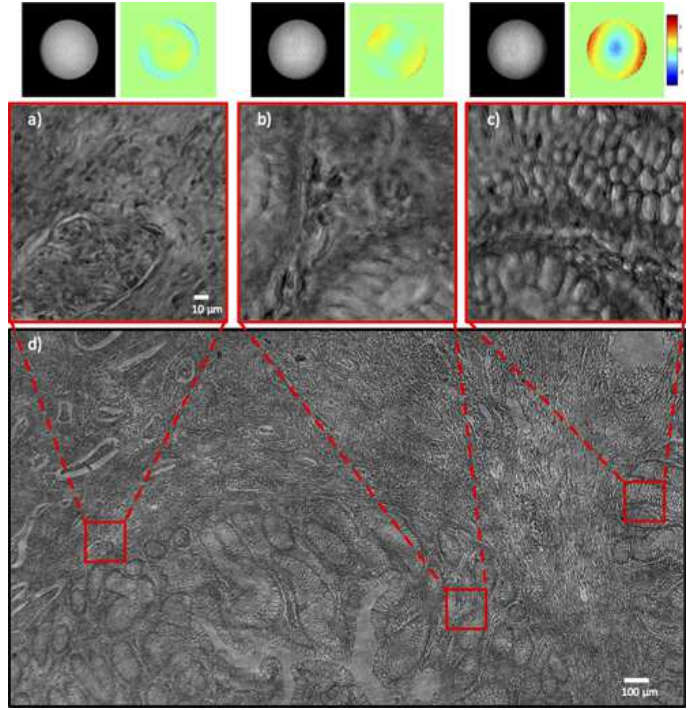}
\caption{\label{fig:histology}Wide FOV histology image. (a)-(c) show FP reconstructed amplitudes of the sub-regions in the full FOV image in (d). Simultaneous to the sample field reconstruction, FP algorithm also characterizes the pupil function's amplitude and phase of each sub-region to reconstruct aberration-free high-resolution images.}
\end{figure}

To demonstrate the wide-field performance of our laser FPM, we capture a full FOV image of an H\&E stained histology sample, as shown in Fig. \ref{fig:histology}. We first segment the entire region into small square tiles (370 ${\rm \mu m}$ by 370 ${\rm \mu m}$) to account for the spatially varying aberration of our imaging system. Then we apply our modified FP algorithm to each tile to reconstruct a high resolution image of the entire FOV. In the end, we are able to correct for the spatially varying aberration and obtain a wide-field and high-resolution image, just like in the original FPM with LEDs as the illumination source but at a much higher capturing speed.

\section{Discussion}
We demonstrate that an FPM setup involving a laser as the illumination source is capable of providing a both wide FOV and high-resolution image. Although the higher temporal and spatial coherence of laser compared to those of an LED lead to speckle artifacts, appropriately including the additional constraint by the DPC-deconvolved phase in our FP algorithm is able to mitigate the negative influences on the final reconstructed image. However, we acknowledge that this modification is not a complete solution to the speckle noise because 1) DPC deconvolution only works for a weakly absorbing and weakly scattering sample; and 2) the background amplitude fluctuation in the captured low-resolution images due to speckles is not directly addressed by the algorithm. In order to significantly reduce the speckles which predominantly originate from the interference of the unscattered laser beam, all the glass surfaces in the optical system would need to have anti-reflective coatings suitable for the laser's wavelength. 

The use of 3D-printed mirror elements allows for intuitive optical setup, but it is not as flexible as other illumination schemes when different objective lenses are used in the imaging system. An entirely new array may be required for different objectives to satisfy the desired resolution gain and appropriate overlap in the Fourier domain. A modular or adjustable design of the array would make the system more flexible in different imaging scenarios. 

Use of Galvo mirrors to direct the collimated laser beam allows for the efficient usage of the laser power, and liberates the illumination source from being the bottleneck of FPM's capturing speed. With a faster camera sensor and an easily adjustable illumination arrangement, imaging faster dynamic samples with FPM will be possible. Moreover, the proposed setup can accommodate lasers with any wavelengths that are compatible with the optical elements and the sensor by simply coupling the laser output to the collimation optics. Bright-field spectral imaging can be easily realized by utilizing a multispectral laser as the illumination source. The fast frame rate and the variety of laser wavelengths available make FPM more attractive for a wider usage.

\section{Appendix A: quantifying resolution using Siemens star target}

\begin{figure}
\centering\includegraphics[width=3cm]{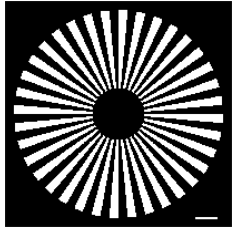}
\caption{\label{fig:star}The Siemens star used for quantifying our system's resolution performance. It consists of 36 line pairs extending radially, which are etched on a gold-coated microscope slide by a focused ion beam. The scale bar is 5 ${\rm \mu m}$}
\end{figure}

The rise of numerous illumination light sources (e.g. LEDs and lasers) and computational image processing algorithms necessitate a need for a better resolution target to quantify an imaging system's resolution than the conventional metrics such as a two-point resolution target \cite{HorstmeyerStar}. The resolution target proposed by Ref. \cite{HorstmeyerStar} is a Siemens star target which is a pattern of spokes, as shown in Fig. \ref{fig:star}. The Siemens star we use is 46 ${\rm \mu m}$ in diameter, with 36 periodic patterns extending in the radial direction. 

\begin{figure} [b]
\centering\includegraphics[width=9cm]{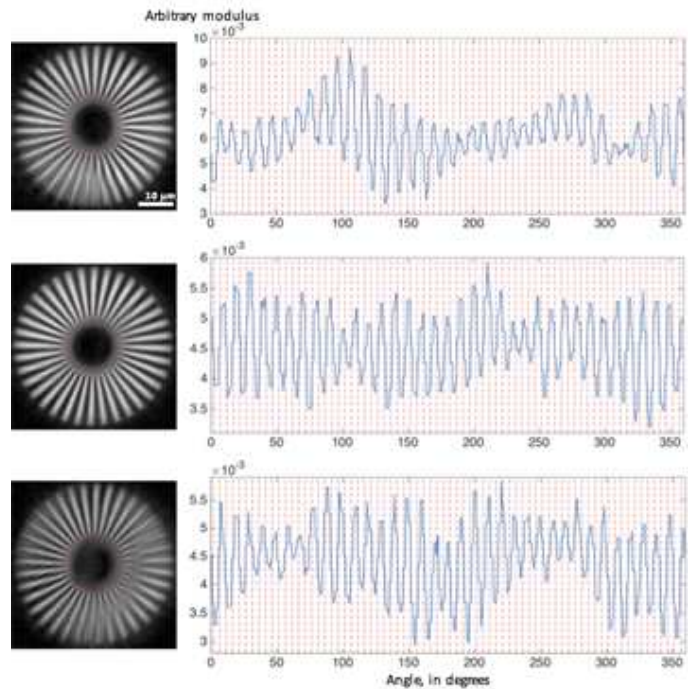}
\caption{\label{fig:starplot}Reconstruction results of Siemens star target placed at the FOV's center, 1 mm away from the center, and 1.13 mm away from the center. The line plot shows that the periodic pattern is successfully resolved along the arc with a radius 8.54 ${\rm \mu m}$. The periodicity at this point is 1.49 ${\rm \mu m}$, which matches our theoretical expectation. }
\end{figure}

The target is translated to three different locations of the FOV: the center, 1 mm away from the center, and 1.13 mm away from the center. The illumination source used in this particular experiment is a He-Ne laser with 632 nm wavelength. In all cases, the smallest observable spoke periodicity occurs at 8.54 ${\rm \mu m}$ radially away from the pattern's origin, with the periodicity being equivalent to 1.49 ${\rm \mu m}$. The line plots along the circumference defined by the observable radius in all three locations are shown in Fig. \ref{fig:starplot}. The observed resolution matches closely with the theoretical resolution given by the wavelength of illumination and the system's NA: periodicity = $\frac{632 {\rm nm}}{0.425} = 1.487 {\rm \mu m}$.

\section{Acknowledgements}
This project was funded by the National Institute of Health (NIH) Agency Award: R01 AI096226; and the Caltech Innovation Initiative (CII): 25570015. We thank Daniel Martin for fabricating the Siemens star target, and Mooseok Jang and Haowen Ruan for helpful discussions. 

\end{document}